\documentclass[epj]{svjour}

\usepackage{graphicx}
\usepackage{fancyhdr}
\def\makeheadbox{{
 }}
\def\mytitle{My title} 
\def\myauthors{My name}  
\def\mytype{My type of session}
\def\mysession{My session}

\def\mytitle{FCNCs in a SO(10) SUSY GUT with Family Symmetry} 
\def\myauthors{Wolfgang Altmannshofer}    
\def\mytype{Contributed Talk}    
\def\mysession{Flavor Physics}

\begin{document}
\title{Flavor Changing Neutral Current Processes in a SO(10) SUSY GUT with Family Symmetry
}
\author{Wolfgang Altmannshofer 
\thanks{\emph{Email:} wolfgang.altmannshofer@ph.tum.de}
}                     
%
%
\institute{
Physik Department, 
Technische Universit\"at M\"unchen,
D-85748 Garching, Germany
}
%
\date{October 8, 2007}
\abstract{
We report on a detailed analysis of a SO(10) SUSY GUT model of Der\-m\'{i}\-\v{s}ek and Raby (DR) with a $D_3$ family symmetry. The model is completely specified in terms of only 24 parameters and is able to successfully describe both quark and lepton masses and mixings, except for $|V_{ub}|$ that turns out to be too low. However, a global fit shows that flavor changing (FC) processes like $B_s \to \mu^+ \mu^-$, $B_s$-mixing, $B^+ \to \tau^+ \nu$, $B \to X_s \gamma$ and $B \to X_s \ell^+ \ell^-$ pose a serious problem to the DR model. The simultaneous description of these FC processes forces squarks to have masses well above 1 TeV, not appealing on grounds of naturalness and probably beyond the reach of the LHC.
} 
\maketitle
%
\section{Introduction}
\label{sec:intro}
Many extensions of the Standard Model (SM), like the general Minimal Supersymmetric Standard Model (MSSM), typically introduce a large set of additional parameters to the SM ones and therefore largely lose their predictivity. On the other hand, in Supersymmetric Grand Unified Theories (SUSY GUTs) with additional family symmetries, the number of parameters can even be smaller than in the SM.
In a top down approach, these models then allow to predict observables at the low scale in terms of a manageable number of GUT scale parameters.

One of such highly predictive models is the SO(10) SUSY GUT introduced by Der\-m\'{i}\-\v{s}ek and Raby in \cite{DR05}. In \cite{DR06} lepton flavor violating processes and electric di\-pole moments were studied extensively within this mo\-del. Here we report on a detailed analysis \cite{AABGS} of the same model in the light of the best measured FC processes in the quark sector.
%
\section{The Model}
\label{sec:model}
The DR model is a supersymmetric SO(10) Grand Unified Theory that is supplemented by a $D_3 \times [ U(1) \times Z_2 \times Z_3]$ family symmetry. 

The three generations of quarks and leptons are each unified in a {\bf 16}. Furthermore, the model has one additional {\bf 10} that contains the two Higgs doublets of the MSSM. The family symmetry ensures that only a universal third generation Yukawa coupling is allowed. Yukawa couplings for the first and second generation are then generated by a Froggatt-Nielsen mechanism \cite{FN}. It turns out that the resulting Yukawa matrices can be parameterized in terms of only 11 parameters. 
All parameters of the model are summarized in Table~\ref{tab:parameters}. Among them, the most important role in the numerical analysis of the FC processes is played by the universal sfermion mass $m_{16}$, the universal trilinear coupling $A_0$ and the Higgsino mass parameter $\mu$.
Moreover, $\tan\beta$ is forced to be around 50, because of third generation Yukawa unification.
\begin{table}
\caption{The 24 parameters in the DR model. (The Yukawa textures $\rho$, $\sigma$, $\tilde \epsilon$ and  $\xi$ are complex.)}
\label{tab:parameters}
\begin{tabular}{lcc}
\hline\noalign{\smallskip}
Sector & \# & Parameters \\
\noalign{\smallskip}\hline\noalign{\smallskip}
Yukawa textures & 11 & $\epsilon$, $\epsilon'$, $\lambda$, $\rho$, $\sigma$, $\tilde \epsilon$, $\xi$ \\
RH neutrinos & 3 & $M_{R_1}$, $M_{R_2}$, $M_{R_3}$ \\
gauge couplings & 3 & $\alpha_G$, $M_G$, $\epsilon_3$ \\
SUSY (GUT scale) & 5 & $M_{1/2}$, $m_{16}$, $A_0$, $m_{H_u}$, $m_{H_d}$ \\
SUSY (EW scale) & 2 & $\tan \beta$, $\mu$ \\
\noalign{\smallskip}\hline
\end{tabular}
\end{table}

The total number of model parameters is 24 and once they are fixed, the complete MSSM Lagrangian at the electro-weak scale is specified.
%
\section{Basic Procedure of the Analysis}
\label{sec:procedure}
Starting with the model parameters at the GUT scale, the Yukawa matrices, the right-handed (RH) neutrino mass matrix, the gauge couplings and the soft SUSY breaking parameters are run down using renormalization group equations. The RH neutrinos are integrated out at their respective scale and the remaining parameters
are further run down to the electro-weak scale, where also the parameters $\mu$ and $\tan\beta$ are specified. The complete set of MSSM parameters is then given in terms of the original model parameters from Table~\ref{tab:parameters}.
This is especially true for the flavor off-diagonal entries of the squark mass matrices that are generated radiatively in the running procedure because of the Yukawa couplings. As the model features no additional sources of flavor violation and in particular no other CP phases than the ones appearing in the CKM and PMNS matrices, it can be classified as being minimal flavor violating \cite{MFV}.

Having at hand all MSSM parameters, one calculates the SUSY spectrum, loop corrections to fermion masses and mixings and finally also FCNC observables like $\Delta M_s$, ${\rm BR}(B_s \to \mu^+ \mu^-)$, ${\rm BR}(B \to X_s \gamma)$ and ${\rm BR}(B \to X_s \ell^+ \ell^-)$ as well as the branching ratio for the FC decay $B^+ \to \tau^+ \nu$.

\begin{figure}
\includegraphics[width=0.45\textwidth]{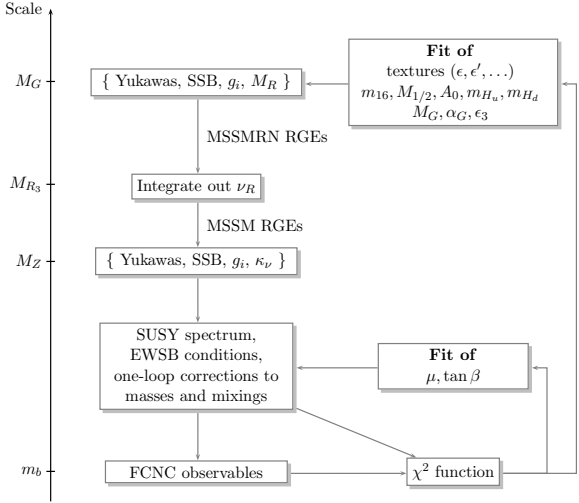}
\caption{Schematic chart of the strategy followed in the fitting procedure of \cite{AABGS}.}
\label{fig:chart}
\end{figure}

Using these FC observables as well as flavor conserving quantities like gauge couplings and fermion masses and mixing angles, a $\chi^2$ function is defined \cite{AABGS}, that is then minimized by varying the original model parameters. (See Fig.~\ref{fig:chart} for a schematic chart of this whole procedure.)

Using only flavor conserving observables, it was shown in \cite{DR05} that the DR model is indeed able to give successfull fits to the experimental data. 
In particular, the model is able to excellently reproduce the CKM matrix elements except for $V_{ub}$, whose absolute value typically comes out as
\begin{equation}\label{VubDR}
|V_{ub}^{\rm DR}| \approx 3.2\times10^{-3}~.
\end{equation}
This number is smaller than both the value from the exclusive and the inclusive determination \cite{UTfit}
\begin{eqnarray}\label{eq:Vub}
|V_{ub}|_{\rm excl}^{\rm exp} &=& (3.50 \pm 0.40)\times 10^{-3}~,\nonumber\\
|V_{ub}|_{\rm incl}^{\rm exp} &=& (4.49 \pm 0.33)\times 10^{-3}~.
\end{eqnarray}
The main novelty of the analysis in \cite{AABGS} was then to also include the abovementioned FC observables into the fit.
%
\section{Interplay of Flavor Changing Processes}
\label{sec:FCNCs}
Although the DR model is minimal flavor violating, one expects interesting effects in various FC processes due to the large value of $\tan\beta$.\footnote{In the numerical analysis, we resum large $\tan\beta$ corrections following \cite{BCRS}.} In this section we discuss the general pattern of these effects.
%
\subsection{\boldmath $B_s \to \mu^+ \mu^-$ and $B_s$ mixing}
\label{sec:Bmumu}
Combining data from CDF and D\O~results in the following upper bound on the branching ratio of the rare decay $B_s \to \mu^+ \mu^-$ at 95\% C.L.
\begin{equation}\label{eq:BsmumuEXP}
{\rm BR}(B_s \to \mu^+ \mu^-)^{\rm exp} < 5.8 \times 10^{-8}~,
\end{equation}
that is still much larger than the SM prediction \cite{BsmumuDeltaMs,AABGS}
\begin{equation}\label{eq:BsmumuSM}
{\rm BR}(B_s \to \mu^+ \mu^-)^{\rm SM} = (3.37 \pm 0.31) \times 10^{-9}~.
\end{equation}
The helicity suppression of the SM result can be lifted in the MSSM with large $\tan\beta$ by neutral Higgs penguins \cite{BsmumuTANB} that lead to contributions to the branching ratio that are strongly enhanced by $\tan\beta$
\begin{equation}\label{eq:BRBsmumu}
{\rm BR}(B_s \to \mu^+ \mu^-) \propto \frac{\tan^6\beta}{M_{A}^4}~.
\end{equation}
The same mechanism also leads to neutral Higgs double penguin contributions to the mass difference in the $B_s - \bar B_s$ system \cite{BCRS}
\begin{equation}\label{eq:DMs}
(\Delta M_s)^{\rm DP} \propto - \frac{\tan^4\beta}{M_{A}^2}~.
\end{equation}
On the experimental side, this quantity is known very precisely \cite{DeltaMsEXP}
\begin{equation}\label{eq:DMsEXP}
(\Delta M_s)^{\rm exp} = (17.77 \pm 0.10 \pm 0.07) {\rm ps}^{-1}~.
\end{equation}
On the other hand, the theory prediction in the SM suffers from large hadronic uncertainties \cite{UTfit}
\begin{equation}\label{eq:DMsSM}
(\Delta M_s)^{\rm SM} = (18.6 \pm 2.3) {\rm ps}^{-1}~,
\end{equation}
leaving still some room for new physics contributions. But as $\tan\beta$ is forced to be around 50 by third generation Yukawa unification, both observables constrain the pseudoscalar Higgs mass $M_A$. In fact, in the DR model we find a lower bound on $M_A > 450~{\rm GeV}$, that then approximately also holds for the other heavy Higgs particles.
%
\subsection{\boldmath $B^+ \to \tau^+ \nu$}
\label{sec:Btaunu}
Using the most recent experimental results one obtains the following average for the branching ratio of the tree level decay $B^+ \to \tau^+ \nu$ (see \cite{AABGS} and references therein)
\begin{equation}\label{eq:BtaunuEXP}
{\rm BR}(B^+ \to \tau^+ \nu)^{{\rm exp}} = (1.41 \pm 0.43) \times 10^{-4}~.
\end{equation}
The SM branching ratio is proportional to $|V_{ub}|^2$. Using the exclusive and the inclusive value for $|V_{ub}|$ from eq.~(\ref{eq:Vub}) yields the following SM predictions \cite{AABGS}
\begin{eqnarray}\label{eq:BtaunuSM}
{\rm BR}(B^+ \to \tau^+ \nu)^{\rm SM}_{\rm excl} &=& (0.80\pm0.20)\times10^{-4}~, \nonumber\\
{\rm BR}(B^+ \to \tau^+ \nu)^{\rm SM}_{\rm incl} &=& (1.31\pm0.23)\times10^{-4}~.
\end{eqnarray}
In the MSSM there is an additional contribution to this decay coming from the exchange of a charged Higgs boson. It interferes destructively with the SM contribution and one finds \cite{RBtaunu,CMW,AABGS}
\begin{eqnarray}\label{eq:RBtaunu}
R_{B\tau\nu} &=& 
\frac{{\rm BR}(B^+ \to \tau^+\nu)^{\rm DR}}{{\rm BR}(B^+ \to \tau^+\nu)^{\rm SM}} = \nonumber\\
&=& \left(1 - \frac{M_{B^+}^2}{M_{H^+}^2} \frac{\tan^2\beta}{1+\epsilon_0 \tan\beta}\right)^2 \left\vert \frac{V_{ub}^{\rm DR}}{V_{ub}^{\rm SM}} \right\vert^2~. 
\end{eqnarray}
As confirmed in \cite{AABGS}, the value for $|V_{ub}|$ in the DR model is always lower than in the SM, which leads to a further suppression of the branching ratio with respect to the SM value. In the DR model, we typically find
\begin{equation}\label{eq:BtaunuDR}
{\rm BR}(B^+ \to \tau^+ \nu)^{{\rm DR}} < 0.6 \times 10^{-4}~,
\end{equation}
which is however not yet excluded, given the large experimental error in eq.~(\ref{eq:BtaunuEXP}).
%
\subsection{\boldmath $B \to X_s \gamma$ and $B \to X_s \ell^+ \ell^-$}
\label{sec:bsgamma}
The experimental value for the branching ratio of the inclusive decay $B \to X_s \gamma$ reads \cite{HFAG}
\begin{equation}\label{eq:BXsgammaEXP}
{\rm BR}(B \to X_s \gamma)^{{\rm exp}} = (3.55 \pm 0.27) \times 10^{-4}~,
\end{equation}
which is slightly above the NNLO SM prediction \cite{Misiak}
\begin{equation}\label{eq:BXsgammaSM}
{\rm BR}(B \to X_s \gamma)^{{\rm SM}} = (3.15 \pm 0.23) \times 10^{-4}~.
\end{equation}
As discussed in Sec.~\ref{sec:Bmumu}, Higgs masses are forced to be quite large, implying that new physics contributions to $C_7$, the Wilson coefficient governing $B \to X_s \gamma$, are dominated by chargino - stop loops. For large values of $\tan\beta$ these chargino contributions obey the following approximate relation \cite{LPV,CMW}
\begin{equation}\label{eq:C7Chargino}
C_7^{\tilde \chi^+} \propto \mu A_t \tan\beta \times {\rm sign}(C_7^{\rm SM})~.
\end{equation}
For $\mu > 0$ and $A_t < 0$ the sign of the chargino contribution is opposite to the SM one. Without invoking further constraints, the model favors very large chargino contributions $C_7^{\tilde \chi^+} \approx -2 C_7^{\rm SM}$ that lead to $C_7 \approx - C_7^{\rm SM}$ which accommodates the data on $B \to X_s \gamma$. 

A further important process to be considered is then $B \to X_s \ell^+ \ell^-$. Both the forward backward asymmetry and the branching ratio of this decay are sensitive to the sign of $C_7$. In case the sign of $C_7$ is opposite to its SM value, the forward backward asymmetry has no zero, which is however not yet excluded experimentally. On the other hand, the experimental data on the branching ratio in the low $s$ region $1~{\rm GeV}^2 < s < 6~{\rm GeV}^2$ \cite{BXsllEXP}
\begin{equation}\label{eq:BXsllEXP}
{\rm BR}(B \to X_s \ell^+ \ell^-)^{{\rm exp}} = (1.60 \pm 0.51) \times 10^{-6}~,
\end{equation}
is in very good agreement with the SM prediction \cite{HLMW}
\begin{equation}\label{eq:BXsllSM}
{\rm BR}(B \to X_s \ell^+ \ell^-)^{{\rm SM}} = (1.59 \pm 0.11) \times 10^{-6}~.
\end{equation}
It has been shown \cite{GHM,LPV} that the experimental result (\ref{eq:BXsllEXP}) excludes the ``wrong sign'' solution for $C_7$ if the Wilson coefficients $C_9$ and $C_{10}$ are SM-like. This is especially the case in a minimal flavor violating MSSM \cite{ALGH} and also in the DR model. Thus chargino contributions to $C_7$ have to be suppressed, which can only be done by raising the stop masses.
%
\section{Results of the Numerical Analysis}
\label{sec:results}
The main features of the interplay of the FC processes described in Sec.~\ref{sec:FCNCs} are then also reflected in the performed numerical fits. The adopted strategy in \cite{AABGS} was to roughly set the scale for the sfermion masses by fixing $m_{16}$, while all the other model parameters where left free in the fits.
%
\subsection{Fits with \boldmath $\mu > 0$}
\label{sec:mupositiv}
\begin{figure}
\includegraphics[width=0.45\textwidth]{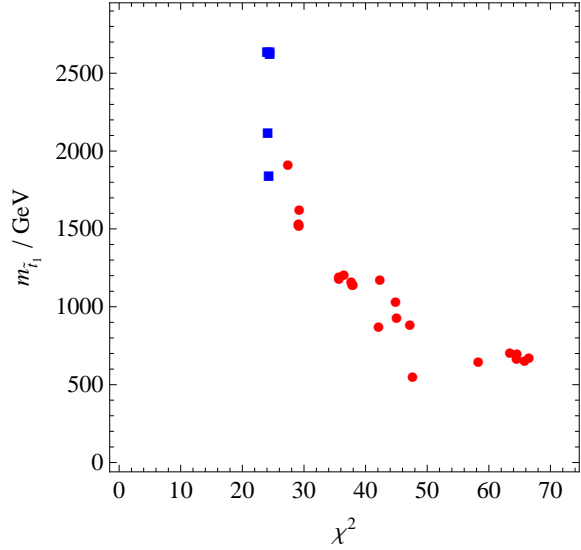}
\caption{Total $\chi^2$ vs. the lightest stop mass for all obtained fits. Red circular points correspond to fits with positive $\mu$, blue squares to negative $\mu$.}
\label{fig:scatter}
\end{figure}
For positive values of $\mu$, the fit strongly prefers values for $A_0$ that obey the following approximate relation at the GUT scale
\begin{equation}\label{eq:A0m16}
A_0 \approx - 2 m_{16}~,
\end{equation}
which helps to obtain third generation Yukawa unification \cite{YukawaUnification} and leads to an inverted mass hierarchy for squarks. These large values for $A_0$ also result in large negative values for $A_t$ at the electro-weak scale, that in turn lead to the large chargino contributions to $B \to X_s \gamma$ discussed in Sec.~\ref{sec:bsgamma}. The only possibility to tame these corrections is then to decouple stops, which can be done by choosing a very large $m_{16}$.

In Fig.~\ref{fig:scatter} the correlation between the lightest stop mass $m_{\tilde t_1}$ and the total $\chi^2$ is shown for all obtained fits. Reasonable fits require stop masses $m_{\tilde t_1} > 1.9~{\rm TeV}$, corresponding to values for $m_{16}$ in the range $(8 - 10)~{\rm TeV}$. For such heavy stops, the chargino corrections to $B \to X_s \gamma$ are under control, but the large stop masses clash with the motivation for SUSY as a solution of the hierarchy problem.

The fit results for some observables in this scenario are collected in Table~\ref{tab:results}. 
%
\subsection{Fits with \boldmath $\mu < 0$}
\label{sec:munegativ}
In \cite{AABGS} also scenarios with negative $\mu$ were considered. In such cases $A_0$ and $m_{16}$ apparently do not have to fullfill relation (\ref{eq:A0m16}).\footnote{Successfull fits away from relation (\ref{eq:A0m16}) were obtained also for positive values of $\mu$. However, these fits perform worse than the corresponding negative $\mu$ fits, because large loop corrections to the bottom quark mass add up constructively and lead to a prediction of $m_b$ that is roughly $4 \sigma$ too large.}
The fit then chooses values for $A_0$ that typically lead to very small $A_t$.
This makes chargino contributions to $C_7$ automatically small (see eq.~(\ref{eq:C7Chargino})), thus solving the problem with $B \to X_s \gamma$. On the other hand, the squark spectrum does not show an inverted hierarchy in this case and the lightest squark (which is usually still a stop) has again a very large mass $m_{\tilde t_1} > 1.8~{\rm TeV}$.

Some fit results for a negative $\mu$ case can again be found in Table~\ref{tab:results}.
\begin{table}
\caption{Fit results for some selected observables for two obtained fits. More detailed tables can be found in \cite{AABGS}.}
\label{tab:results}
\begin{tabular}{lcc}
\noalign{\centering $m_{16} = 10~{\rm TeV}$, $\mu = 1.2~{\rm TeV} $}
\noalign{\smallskip}\hline\noalign{\smallskip}
Observable & Fit value & Pull ($\sigma$)\\
\noalign{\smallskip}\hline\noalign{\smallskip}
$m_{\tilde t_1}~[{\rm TeV}]$ & $1.9$ & --- \\
\noalign{\smallskip}
${\rm BR}(B_s \to \mu^+ \mu^-)\times10^8$ & 2.1 & --- \\
${\rm BR}(B^+ \to \tau^+ \nu)\times10^4$ & 0.517 & 2.1 \\
${\rm BR}(B \to X_s \gamma)\times10^4$ & 2.86 & 1.3 \\
\noalign{\smallskip}\hline
\end{tabular}
\begin{tabular}{lcc}
\noalign{\bigskip}
\noalign{\centering $m_{16} = 4~{\rm TeV}$, $\mu = -2.1~{\rm TeV}$}
\noalign{\smallskip}\hline\noalign{\smallskip}
Observable & Fit value & Pull ($\sigma$)\\
\noalign{\smallskip}\hline\noalign{\smallskip}
$m_{\tilde t_1}~[{\rm TeV}]$ & $2.6$ & --- \\
\noalign{\smallskip}
${\rm BR}(B_s \to \mu^+ \mu^-)\times10^8$ & 0.33 & --- \\
${\rm BR}(B^+ \to \tau^+ \nu)\times10^4$ & 0.59 & 1.9 \\
${\rm BR}(B \to X_s \gamma)\times10^4$ & 3.34 & 0.4 \\
\noalign{\smallskip}\hline
\end{tabular}
\end{table}

\section{Conclusions}
\label{sec:conclusions}
The SO(10) SUSY GUT model of Der\-m\'{i}\-\v{s}ek and Raby \cite{DR05} is able to successfully fit the known quark and lepton masses as well as the CKM and PMNS mixing matrices. The only exception is the absolute value of the CKM matrix element $V_{ub}$ that is even smaller than the central exclusive value.

Given such a small value of $|V_{ub}|$, we then find a very low upper bound (\ref{eq:BtaunuDR}) on the branching ratio of $B^+ \to \tau^+ \nu$ in the DR model. Consequently, this decay will turn out to be quite problematic for the model, if the central experimental value for the branching ratio stays above $1.0 \times 10^{-4}$.

Furthermore, we find that the model is not able to simultaneously fit the branching ratios of the decays $B_s \to \mu^+ \mu^-$, $B \to X_s \gamma$ and $B \to X_s \ell^+ \ell^-$, unless the squark spectrum is made very heavy ($m_{\tilde t_1} > 1.8~{\rm TeV}$). Such large squark masses may be problematic from the point of view of naturalness and squarks may be even beyond the reach of the LHC.

As the example of the DR model shows, it is essential to check simultaneously many flavor changing processes to test the validity of models for fermion masses and mixings.
 
\subsection*{Acknowledgments}
I warmly thank the other authors of \cite{AABGS} for many useful discussions. The work of W.A. is supported by the German Bundesministerium f{\"u}r Bildung und Forschung under contract 05HT6WOA.

\end{document}